\documentclass[preliminary]{eptcs}
\providecommand{\event}{WORDS 2011} 
\usepackage{breakurl}             
\usepackage{epsfig}
\usepackage{latexsym}
\usepackage{amssymb}
\usepackage{amsfonts}
\usepackage{amsmath}
\newtheorem{definition}{Definition}[section]
\providecommand{\urlalt}[2]{\href{#1}{#2}}
\providecommand{\doi}[1]{doi:\urlalt{http://dx.doi.org/#1}{#1}}

\title{Pattern $1^j 0^i$ avoiding binary words}
\author{Stefano Bilotta \qquad\qquad Elisa Pergola \qquad\qquad
Renzo Pinzani\\ \institute{Dipartimento di Sistemi e Informatica\\
Universit\`a degli Studi di Firenze\\ Viale
 G.B. Morgagni 65, 50134 Firenze, Italy}\\
\email{bilotta@dsi.unifi.it \qquad\quad elisa@dsi.unifi.it
\qquad\quad pinzani@dsi.unifi.it}}
\def\titlerunning{Pattern $1^j 0^i$ avoiding binary words}
\def\authorrunning{S. Bilotta, E. Pergola \& R. Pinzani}

\begin{document}

\maketitle

\begin{abstract}
In this paper we study the enumeration and the construction,
according to the number of ones, of particular binary words
avoiding the fixed pattern $\mathfrak{p}(j,i)=1^j 0^i$, $0 < i <
j$. The growth of such words can be described by particular
jumping and marked succession rules. This approach enables us to
obtain an algorithm which constructs all binary words having a
fixed number of ones and then kills those containing the forbidden
pattern.
\end{abstract}

\section{Introduction}
The problem of determining the appearance of a fixed
\emph{pattern} in long sequences of observation is interesting in
many scientific problems.

For example in the area of computer network security, intrusions
are becoming increasingly frequent and their detection is very
important. Intrusion detection is primarily concerned with the
detection of illegal activities and acquisitions of privileges
that cannot be detected with information flow and access control
models. There are several approaches to intrusion detection, but
recently this subject has been studied in relation to pattern
matching (see \cite{1,9,12}).

In the area of computational biology, for example, it could be
interesting to detect the occurrences of a particular pattern in a
genomic sequence over the alphabet $\{A,G,C,T\}$, see for instance
\cite{16,18}.

These kinds of applications are interested in the study concerning
both the enumeration and the construction of particular words
avoiding a given pattern over an alphabet $\Sigma$.

In particular, binary words avoiding a fixed pattern
$\mathfrak{p}=p_0...p_{h-1}\in \{0,1\}^h$ constitute a regular
language and can be enumerated in terms of the number of bits $1$
and $0$ by using classical results (see, e.g., \cite{10,11,17}).
Recently, in \cite{2,13}, this subject has been studied in
relation to the theory of Riordan arrays.

In \cite{5}, the authors study the enumeration and the
construction, according to the number of ones, of the class
$F^{[\mathfrak{p}(j)]}$, that is, the class $F \subset \{0,1\}^*$ of
binary words $w$ excluding the fixed pattern
$\mathfrak{p}(j)=1^{j+1}0^j$, $j \geq 1$, such that $|w|_0 \leq
|w|_1$ for any $w \in F$, $|w|_0$ and $|w|_1$ being the number of
zeroes and ones in the word $w$, respectively. The enumeration
problem, according to the number of ones, is solved algebraically
by means of Riordan arrays theory. This approach gives a
\emph{jumping and marked succession rule} describing the growth of
such words. Moreover, in \cite{5} was introduced an algorithm for
constructing all binary words having a fixed number of ones and
excluding those containing the forbidden pattern
$\mathfrak{p}(j)=1^{j+1}0^j$, $j \geq 1$.

In this paper, we focus on the generalization of the fixed
forbidden pattern $\mathfrak{p}$, passing from
$\mathfrak{p}(j)=1^{j+1}0^j$, $j \geq 1$ to
$\mathfrak{p}(j,i)=1^j0^i$, $0<i<j$.

In this case the theory of Riordan arrays is not applicable, while
it is possible to adapt the succession rule for the class
$F^{[\mathfrak{p}(j)]}$ with $\mathfrak{p}(j)=1^{j+1}0^j$, $j \geq
1$, to the class $F^{[\mathfrak{p}(j,i)]}$ for any
$\mathfrak{p}(j,i)=1^j0^i$, $0<i<j$.

The paper is organized as follows. In Section 2 we give some basic
definitions and notation related to the notions of succession rule
and generating tree. In particular, we introduce the concept of
\emph{jumping and marked succession rules} (see \cite{7,8}) which
are succession rules acting on the combinatorial objects of a
class and producing sons at different levels where appear marked
or non-marked labels.

In Section 3, we give a construction, according to the number of
ones, for the set $F^{[\mathfrak{p}(j,i)]}$ for any fixed
forbidden pattern $\mathfrak{p}(j,i)=1^j0^i$, $0<i<j$, by means of
particular jumping and marked succession rules related to the form
of the words in $F$.

\section{Basic definitions and notations}
A \emph{succession rule} $\Omega$ is a system constituted by an
\emph{axiom} $(a)$, with $a \in \mathbb{N}$, and a set of
\emph{productions} of the form:
\begin{displaymath}
(k)\rightsquigarrow (e_1(k))(e_2(k))\ldots(e_k(k)), \ \ \ \ \ k
\in \mathbb{N}, \ e_i : \mathbb{N} \rightarrow \mathbb{N}.
\end{displaymath}

A production constructs, for any given label $(k)$, its
\emph{successors} $(e_1(k)),(e_2(k)),\ldots,(e_k(k))$. In most of
the cases, for a succession rule $\Omega$, we use the compact
notation:
\begin{equation}
\label{uno} \left\{
\begin{array}{cl}
 (a) & \\
 (k) & \rightsquigarrow (e_1(k))(e_2(k))\ldots(e_k(k))
\end{array}
\right.
\end{equation}

The rule $\Omega$ can be represented by means of a
\emph{generating tree}, that is a rooted tree whose vertices are
the labels of $\Omega$; where $(a)$ is the label of the root and
each node labelled $(k)$ has $k$ sons labelled
$(e_1(k)),(e_2(k)),\ldots,(e_k(k))$, respectively. As usual, the
root lies at level 0, and a node lies at level $n$ if its parent
lies at level $n-1$. If a succession rule describes the growth of
a class of combinatorial objects, then a given object can be coded
by the sequence of labels met from the root of the generating tree
to the object itself. We refer to \cite{3} for further details and
examples.

The concept of a succession rule was introduced in \cite{6} by
Chung et al. to study reduced Baxter permutations, and was later
applied to the enumeration of permutations with forbidden
subsequences (for details see \cite{4,19}).

We remark that, from the above definition, a node labelled $(k)$
has precisely $k$ sons. A succession rule having this property is
said to be \emph{consistent}. However, we can also consider
succession rules, introduced in \cite{7}, in which the value of a
label does not necessarily represent the number of its sons, and
this will be frequently done in the sequel.

Regular succession rules are not sufficient to handle all the
enumeration problems and so we consider a slight generalization
called \emph{jumping succession rule} \cite{8}. Roughly speaking,
the idea is to consider a set of productions acting on the objects
of a class and producing sons at different levels.

The usual notation to indicate a jumping succession rule is the
following:
\begin{equation}
\label{due} \left\{
\begin{array}{cl}
 (a) & \\
 (k) & \stackrel{1}{\rightsquigarrow} (e_1(k))(e_2(k))\ldots(e_k(k))\\
 (k) & \stackrel{j}{\rightsquigarrow} (d_1(k))(d_2(k))\ldots(d_k(k))
\end{array}
\right.
\end{equation}

The generating tree associated with (\ref{due}) has the property
that each node labelled $(k)$ lying at level $n$ has two sets of
sons, the first set at level $n+1$ and having labels
$(e_1(k)),(e_2(k)),\ldots,(e_k(k))$ and the second one at level
$n+j$, with $j>1$, and having labels
$(d_1(k)),(d_2(k)),\ldots,(d_k(k))$, respectively.

Another generalization is introduced in \cite{14}, where the
authors deal with \emph{marked succession rules}. In this case the
labels appearing in a succession rule can be marked or not,
therefore \emph{marked} labels are considered together with usual
ones. In this way a generating tree can support negative values if
we consider a node labelled $(\overline{k})$ as opposed to a node
labelled $(k)$ lying on the same level.

A \emph{marked generating tree} is a rooted labelled tree where
appear marked or non-marked labels according to the corresponding
succession rule. The main property is that, on the same level,
marked labels kill or annihilate the non-marked ones with the same
label value, in particular the enumeration of the combinatorial
objects in a class is given by the difference between the number
of non-marked and marked labels lying on a given level.

For any label $(k)$, we introduce the following notation for
generating tree specifications:
\begin{itemize}
\item[] $(\overline{\overline{k}})=(k)$; \item[] $(k)^n =
\underbrace{(k)\ldots(k)}_{n}, \ \ n > 0.$
\end{itemize}

Each succession rule (\ref{uno}) can be trivially rewritten as
(\ref{tre})
\begin{equation}
\label{tre} \left\{
\begin{array}{ll}
 (a) & \\
 (k) & \rightsquigarrow (e_1(k))(e_2(k))\ldots(e_k(k))(k)\\
 (k) & \rightsquigarrow (\overline{k})
\end{array}
\right.
\end{equation}

\begin{figure}[htb]
\begin{center}
\epsfig{file=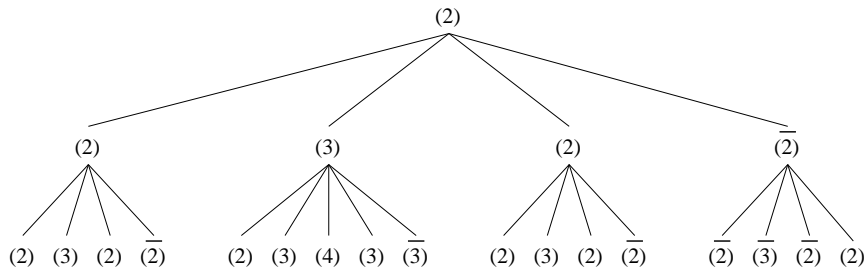,width=4.5in,clip=} \caption{\small{Three
levels of the generating tree associated with the succession rule
(\ref{quattro})} \label{exmar}}\vspace{-15pt}
\end{center}
\end{figure}

For example, the classical succession rule for Catalan numbers can
be rewritten in the form (\ref{quattro}) and Figure \ref{exmar}
shows some levels of the associated generating tree.
\begin{equation}
\label{quattro} \left\{
\begin{array}{ll}
 (2) & \\
 (k) & \rightsquigarrow (2)(3)\ldots(k)(k+1)(k)\\
 (k) & \rightsquigarrow(\overline{k})
\end{array}
\right.
\end{equation}

The concept of marked labels has been implicity used for the first
time in \cite{15}, then in \cite{7} in relation with the
introduction of the signed ECO-systems. In Section 3, we show how
marked succession rules appear in the enumeration of a class of
particular binary words according to the number of ones. Let $F
\subset \{0,1\}^*$ be the class of binary words $w$ such that
$|w|_0 \leq |w|_1$ for any $w \in F$, $|w|_0$ and $|w|_1$ being
the number of zeroes and ones in $w$, respectively.

In this paper we are interested in studying the subclass
$F^{[\mathfrak{p}]} \subset F$ of binary words excluding a given
pattern $\mathfrak{p}=p_0 \ldots p_{h-1} \in \{0,1\}^h$, i.e. the
word $w \in F^{[\mathfrak{p}]}$ that does not admit a sequence of
consecutive indices $i,i+1,\ldots,i+h-1$ such that $w_i w_{i+1}
\ldots w_{i+h-1} = p_0 p_1 \ldots p_{h-1}$. Each word $w \in F$
can be naturally represented as a lattice path on the Cartesian
plane by associating a \emph{rise step}, defined by $(1,1)$ and
denoted by $x$, to each 1's in $F$, and a \emph{fall step},
defined by $(1,-1)$ and denoted by $\overline{x}$, to each 0's in
$F$. From now on, we refer interchangeably to words or their
graphical representations on the Cartesian plane, that is paths.

\section{A construction for the class $F^{[\mathfrak{p}(j,i)]}$}
In this section, we study the enumeration and the construction for
the set $F^{[\mathfrak{p}(j,i)]}$, where
$\mathfrak{p}(j,i)=x^{j}\overline{x}^i=1^{j}0^i$, $0<i<j$, by
setting jumping and marked succession rules describing the growth
of the set. The succession rules, according to the number of rise
steps or equivalently the number of ones, are related to the form
of the lattice paths in $F$.

First of all, we define a \emph{marked forbidden pattern}
$\mathfrak{p}(j,i)$ as a pattern
$\mathfrak{p}(j,i)=x^{j}\overline{x}^i$, $0<i<j$, whose steps
cannot be divided, they must lie always in that defined sequence.
Therefore, a cut operation is not possible within a marked
forbidden pattern.

We denote a marked forbidden pattern by marking its peak. We say
that a point is strictly contained in a marked forbidden pattern
if it is between two steps of the pattern itself.

\begin{figure}[!htb]
\begin{center}
\epsfig{file=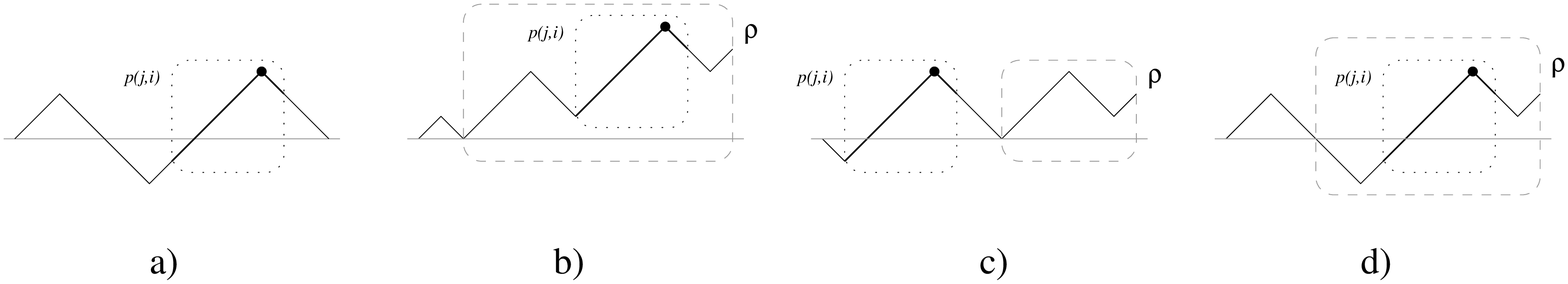,width=6.35in,clip=}
\caption{\small{Some examples of paths in $F$}
\label{def}}\vspace{-15pt}
\end{center}
\end{figure}

In order to study the enumeration and the construction for the
class $F^{[\mathfrak{p}(j,i)]}$, we have to distinguish two cases
depending on the form of the paths in $F$.
\begin{definition}
A path $\omega$ in $F$ is a \emph{$\Delta$-path} if:
\begin{itemize}
\item[$\bullet$] it ends on the $x$-axis (see Figure
\ref{def}.a));

\item[$\bullet$] the ordinate of its endpoint is greater than 0
and its rightmost suffix $\rho$ begins from the $x$-axis by a rise
step and strictly remains above the $x$-axis itself. The suffix
$\rho$ can contain marked forbidden patterns $\mathfrak{p}(j,i)$
(see Figure \ref{def}.b)) or not (see Figure \ref{def}.c)). If
$\rho$ contains marked forbidden patterns $\mathfrak{p}(j,i)$,
then their marked points have ordinate $b \geq j$.
\end{itemize}
\end{definition}

\begin{definition}
A path $\omega$ in $F$ is a \emph{$\Gamma$-path} if the ordinate
of its endpoint is greater that 0 and its rightmost suffix $\rho$
begins from the $x$-axis by a fall step and contains a marked
forbidden pattern $\mathfrak{p}(j,i)$ with ordinate~$b$, $i<b<j$
(see Figure \ref{def}.d)).
\end{definition}

\subsection{$\Delta$-paths in $F$}
For each $\Delta$-path $\omega$ in $F$ having $k$ as the ordinate
of its endpoint, we apply the succession rule (\ref{regola}), for
each $k \geq 0$:

\begin{equation} \label{regola} \left\{
\begin{array}{cl}
 (0) & \\
 (k) & \stackrel{1}{\rightsquigarrow} (0)^2 (1)(2) \cdots (k)(k+1)\\
 (k) & \stackrel{j}{\rightsquigarrow} (\overline{0})^{j-i+1-a}
 (\overline{1})^{j-i-a}(\overline{2})^{j-i-1-a}\ldots(\overline{j-i-1-a})^2(\overline{j-i-a})\dots(\overline{k+j-i})
\end{array}
\right.
\end{equation}

In the second production of (\ref{regola}), the parameter $a$,
with $0 \leq a \leq j-i-1$, is related to the form of the
$\Delta$-path $\omega$ and the way to set $a$ will be described
later in this section.

At this point, we define an algorithm which associates a
$\Delta$-path in $F$ to a sequence of labels obtained by means of
the succession rule (\ref{regola}).

\begin{figure}[!htb]
\begin{center}
\epsfig{file=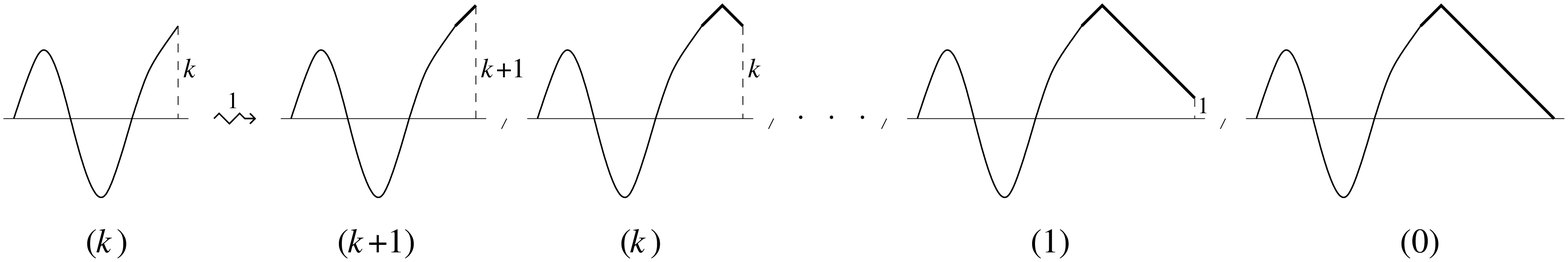,width=5.5in,clip=} \caption{\small{The
mapping associated to $(k) \stackrel{1}{\rightsquigarrow}
(0)(1)(2)\ldots(k+1)$ of (\ref{regola})} \label{sopra}}
\end{center}\vspace{-15pt}
\end{figure}

The axiom $(0)$ is
associated to the empty path $\varepsilon$.\\
A $\Delta$-path $\omega \in F$, with $n$ rise steps and such that
its endpoint has ordinate $k$, provides $k+3$ lattice paths, with
$n+1$ rise steps, according to the first production of
(\ref{regola}) having $0,0,1,\ldots,k+1$ as endpoint ordinate,
respectively. The last $k+2$ labels are obtained by adding to
$\omega$ a sequence of steps consisting of one rise step followed
by $k+1-y$, $0 \leq y \leq k+1$, fall steps (see Figure
\ref{sopra}).

Each lattice path so obtained has the property that its rightmost
suffix beginning from the $x$-axis, either remains strictly above
the $x$-axis itself or ends on the $x$-axis by a fall step. Note
that in this way, the paths ending on the $x$-axis and having a
rise step as last step are never obtained. These paths are bound
to the first label $(0)$ of the first production in (\ref{regola})
and the way to obtain them will be described later in this
section.

\begin{figure}[htb]
\begin{center}
\epsfig{file=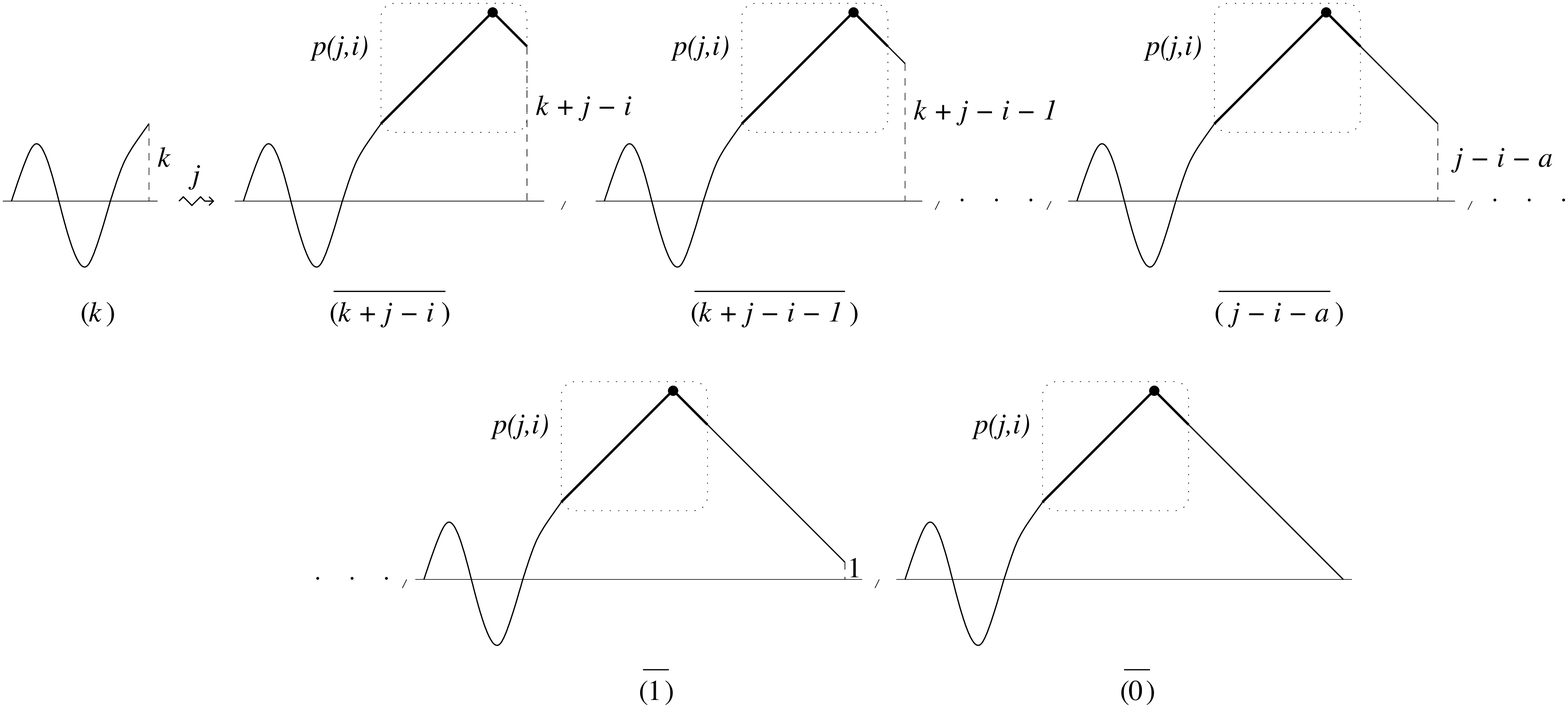,width=5.5in,clip=}
\caption{\small{The mapping associated to $(k)
\stackrel{j}{\rightsquigarrow} (\overline{0})
 (\overline{1})(\overline{2})\ldots(\overline{j-i-a})\dots(\overline{k+j-i})$ of (\ref{regola})}
\label{soprasegnato}}\vspace{-15pt}
\end{center}
\end{figure}

Let the parameter $a$ be fixed, a $\Delta$-path $\omega \in F$,
with $n$ rise steps and such that its endpoint has ordinate $k$,
provides $1+k+j-i+ \sum_{m=0}^{j-i-a-1}{j-i-a-m}$ lattice paths,
with $n+j$ rise steps, according to the second production of
(\ref{regola}) such that $1+k+j-i$ lattice paths having
$0,1,2,\dots,j-i-a,\ldots,k+j-i$ as endpoint ordinate,
respectively, and $j-i-a-m$ lattice paths having $m$ as endpoint
ordinate, $0 \leq m \leq j-i-a-1$. The first $1+k+j-i$ lattice
paths are obtained by adding to $\omega$ a sequence of steps
consisting of the marked forbidden pattern
$\mathfrak{p}(j,i)=x^{j}\overline{x}^i$ followed by $k+j-i-y$ fall
steps, $0 \leq y \leq k+j-i$ (see Figure \ref{soprasegnato}).

Each lattice path so obtained has the property that its rightmost
suffix beginning from the $x$-axis, either remains strictly above
the $x$-axis itself or ends on the $x$-axis by a fall step. At
this point the first label $(0)$ according to the first production
of (\ref{regola}) and the $j-i-a-m$ labels $(\overline{m})$, $0
\leq m \leq j-i-a-1$, according to the second production of
(\ref{regola}), must give lattice paths which either do not
contain marked forbidden pattern in its rightmost suffix and end
on the $x$-axis by a rise step or having the rightmost marked
point with ordinate less than $j$.

In order to obtain the first label $(0)$ according to the first
production of (\ref{regola}), we consider the lattice path
$\omega'$ obtained from $\omega$ by adding a sequence of steps
consisting of one rise step followed by $k$ fall steps. By
applying the previous actions, a path $\omega'$ can be written as
$\omega'= v \varphi$, where $\varphi$ is the rightmost suffix in
$\omega'$ beginning from the $x$-axis and strictly remaining above
the $x$-axis.

We distinguish two cases: in the first one $\varphi$ does not
contain any marked point and in the second one $\varphi$ contains
at least one marked point.

\begin{figure}[htb]
\begin{center}
\epsfig{file=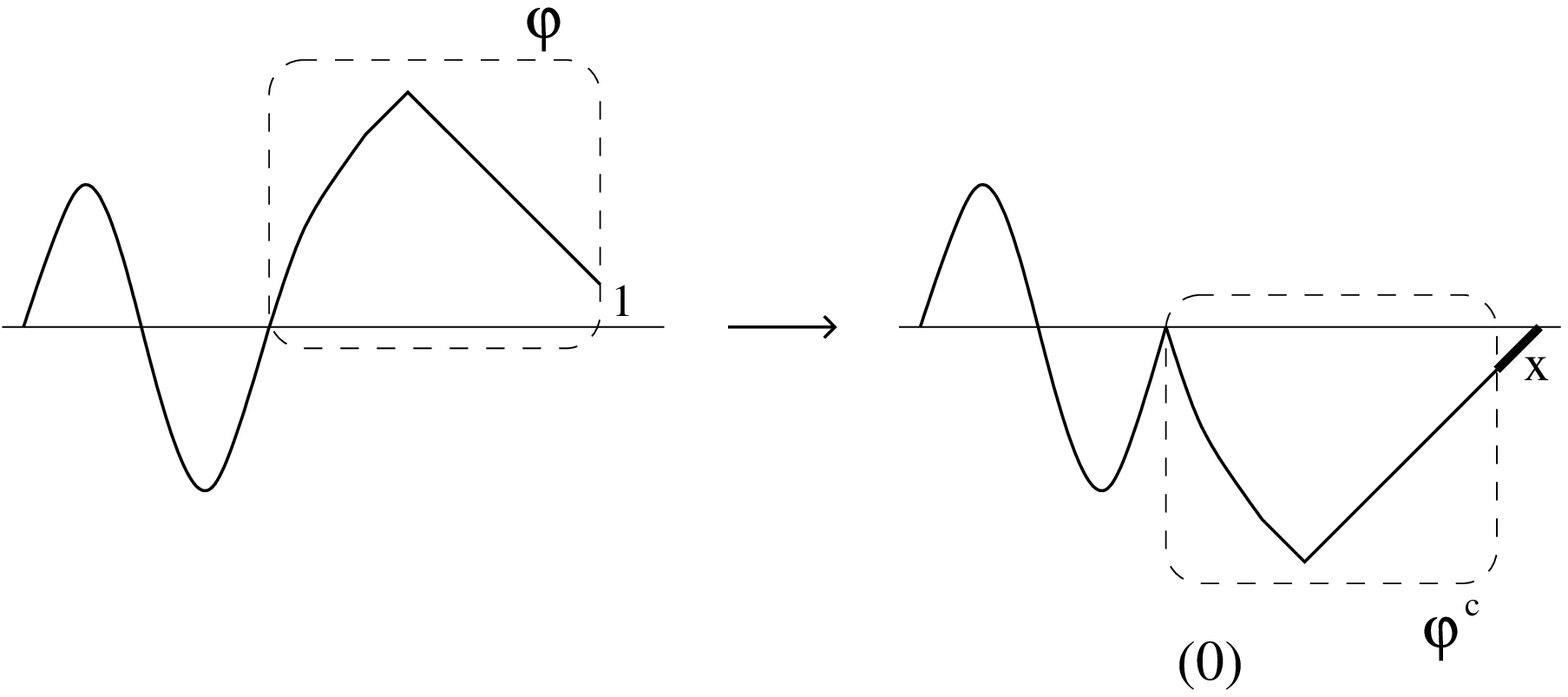,width=3in,clip=} \caption{\small{A
graphical representation of the actions giving the first label
$(0)$ in case of no marked points in $\varphi$}
\label{semplice}}\vspace{-15pt}
\end{center}
\end{figure}

If the suffix $\varphi$ does not contain any marked point, then
the desired label $(0)$ is associated to the path $v\varphi^cx$,
where ${\varphi}^c$ is the path obtained from $\varphi$ by
switching rise and fall steps (see Figure~\ref{semplice}).

\begin{figure}[!htb]
\begin{center}
\epsfig{file=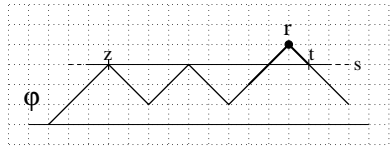,width=2in,clip=} \caption{\small{Marked
point in the suffix $\varphi$: an example with the pattern
$\mathfrak{p}(j,i)=x^2\overline{x}$}\label{ese}}\vspace{-15pt}
\end{center}
\end{figure}

If the suffix $\varphi$ contains marked points, let $r$ be the
rightmost and highest marked point in $\varphi$ and let $t$ be the
nearest and highest point on the right of the marked forbidden
pattern containing $r$ not being strictly within a marked
forbidden pattern. We then consider the straight line $s$ through
the point $t$ and the leftmost and highest point $z$ in $\varphi$
lying above or on the line $s$ and which is not strictly within a
marked forbidden pattern (see Figure \ref{ese}). Obviously if the
straight line $s$ does not intersect any point on the left of $t$
or intersects only points lying strictly within a marked forbidden
pattern, then $z \equiv t$.

\begin{figure}[!htb]
\begin{center}
\epsfig{file=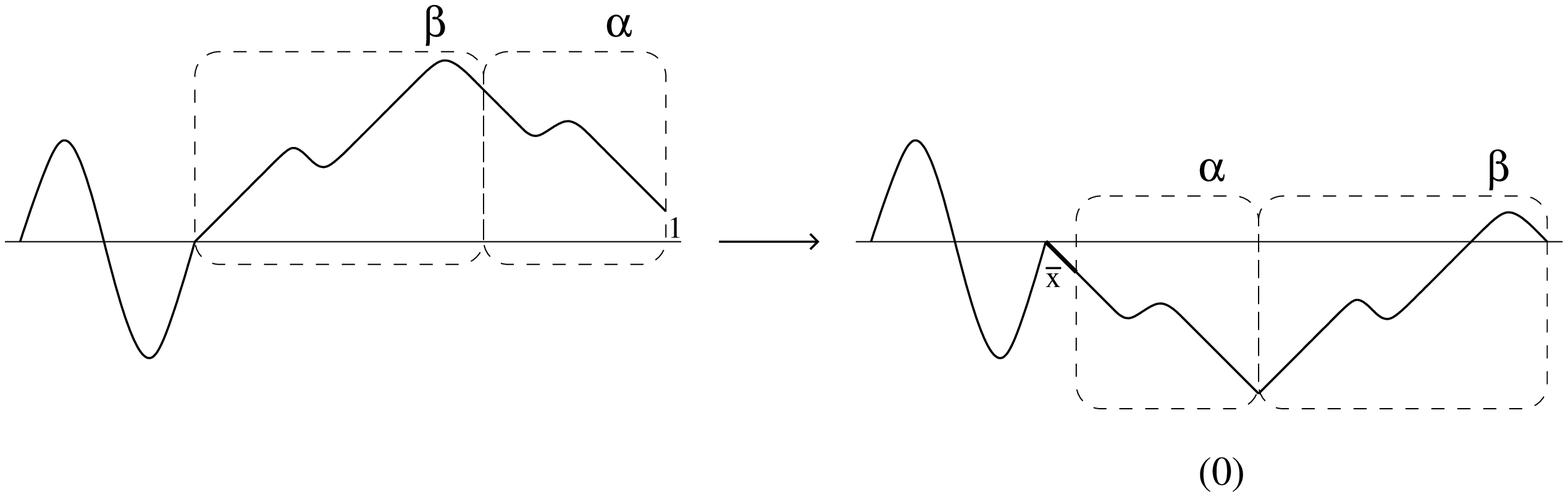,width=3.5in,clip=}
\caption{\small{A graphical representation of the cut and paste
actions giving the first label $(0)$ in case of marked points in
$\varphi$}\label{ardita}}\vspace{-15pt}
\end{center}
\end{figure}

The desired label $(0)$ is associated to the path obtained by
applying the cut and paste actions which consist on the
concatenation of a fall step $\overline{x}$ with the path in
$\varphi$ running from $z$ to the endpoint of the path, say
$\alpha$, and the path running from the initial point in $\varphi$
to $z$, say $\beta$ (see Figure~\ref{ardita}).

In order to obtain the $j-i-a-m$ labels $(\overline{m})$, $0 \leq
m \leq j-i-a-1$, according to the second production of
(\ref{regola}), we consider the paths $\omega''$ obtained from
$\omega$ by adding a sequence of steps consisting of the marked
forbidden pattern $\mathfrak{p}(j,i)=x^{j}\overline{x}^i$ followed
by $y$ fall steps, $k+a \leq y \leq k+j-i-1$. Therefore, we
consider the just obtained paths labelled with
$(\overline{k+j-i-y})$, $k+a \leq y \leq k+j-i-1$, which are
represented in Figure \ref{soprasegnato}.

By performing on each $\omega''$ the cut and paste actions, we
obtain $j-i-a$ paths labelled with

\begin{figure}[!htb]
\begin{center}
\epsfig{file=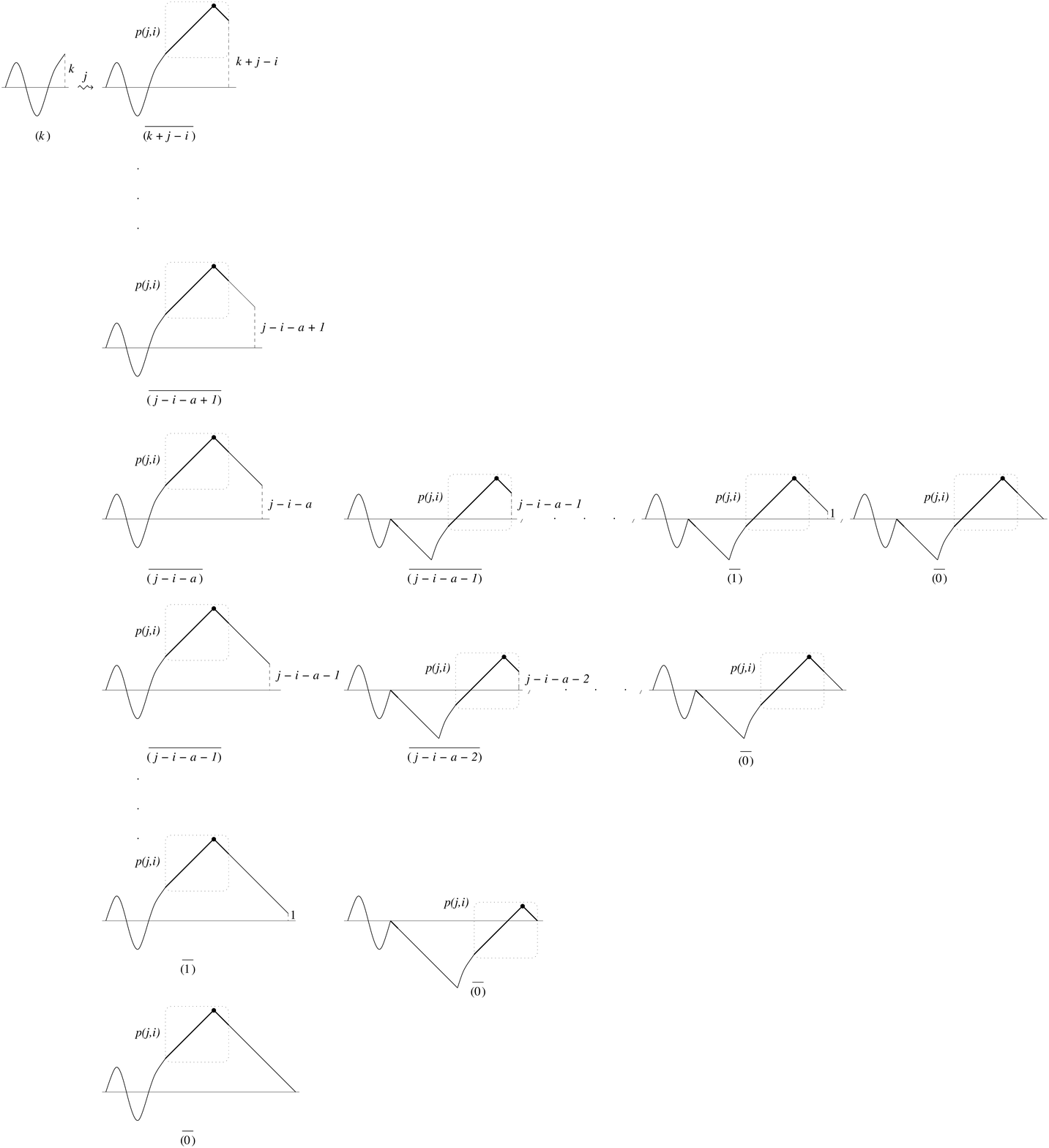,width=6in,clip=}
\caption{\small{The mapping associated to \newline $(k)
\stackrel{j}{\rightsquigarrow} (\overline{0})^{j-i+1-a}
 (\overline{1})^{j-i-a}(\overline{2})^{j-i-1-a}\ldots
 (\overline{j-i-1-a})^2(\overline{j-i-a})\dots(\overline{k+j-i})$ of
(\ref{regola})}\label{global}}\vspace{-15pt}
\end{center}
\end{figure}

\noindent $(\overline{k+j-i-y-1})$, $k+a
\leq y \leq k+j-i-1$. By adding $g$ fall steps, $0 < g \leq
k+j-i-y-1$, to each of such paths (see Figure \ref{global}), we
obtain the complete mapping associated with the second production
of (\ref{regola}).

Note that, we apply the cut and paste actions to the paths
$\omega''$ exclusively. Indeed, by performing the cut and paste
actions to the paths obtained from $\omega$ by adding a sequence
of steps consisting of the marked forbidden pattern
$\mathfrak{p}(j,i)=x^{j}\overline{x}^i$ followed by $y$ fall
steps, $0 \leq y < k+a$, we have repeated paths.\\

In the following is explained the way to set the parameter $a$
related to the form of the $\Delta$-path $\omega$ in $F$ having
$k$ as ordinate of its endpoint.
\begin{itemize}
\item[$\bullet$] If the $\Delta$-path $\omega$ in $F$ has the
ordinate $k$ of its endpoint equal to 0 (or equivalently ends on
the $x$-axis), then $a=0$ and we apply to the path $\omega$ the
production (\ref{propriozero}) for the second production of
(\ref{regola}).

\begin{equation}
\label{propriozero} (0) \stackrel{j}{\rightsquigarrow}
(\overline{0})^{j-i+1}
(\overline{1})^{j-i}(\overline{2})^{j-i-1}\ldots(\overline{j-i-1})^2(\overline{j-i})
\end{equation}

We can observe that, for $k=0$, the paths $\omega''$ related to
the previous construction are the paths obtained from $\omega$ by
adding a sequence of steps consisting of the marked forbidden
pattern $\mathfrak{p}(j,i)=x^{j}\overline{x}^i$ followed by $y$
fall steps, for any $y$ with $0 \leq y \leq j-i-1$. In this case
the point $z$ for the cut and paste actions is always the endpoint
of the path $\omega x^j \overline{x}^i$ having ordinate $j-i$.
Figure \ref{zerata} shows the complete mapping associated to
(\ref{propriozero}) on an example with the pattern
$\mathfrak{p}(j,i)=x^{5}\overline{x}^2$.

\begin{figure}[!htb]
\begin{center}
\epsfig{file=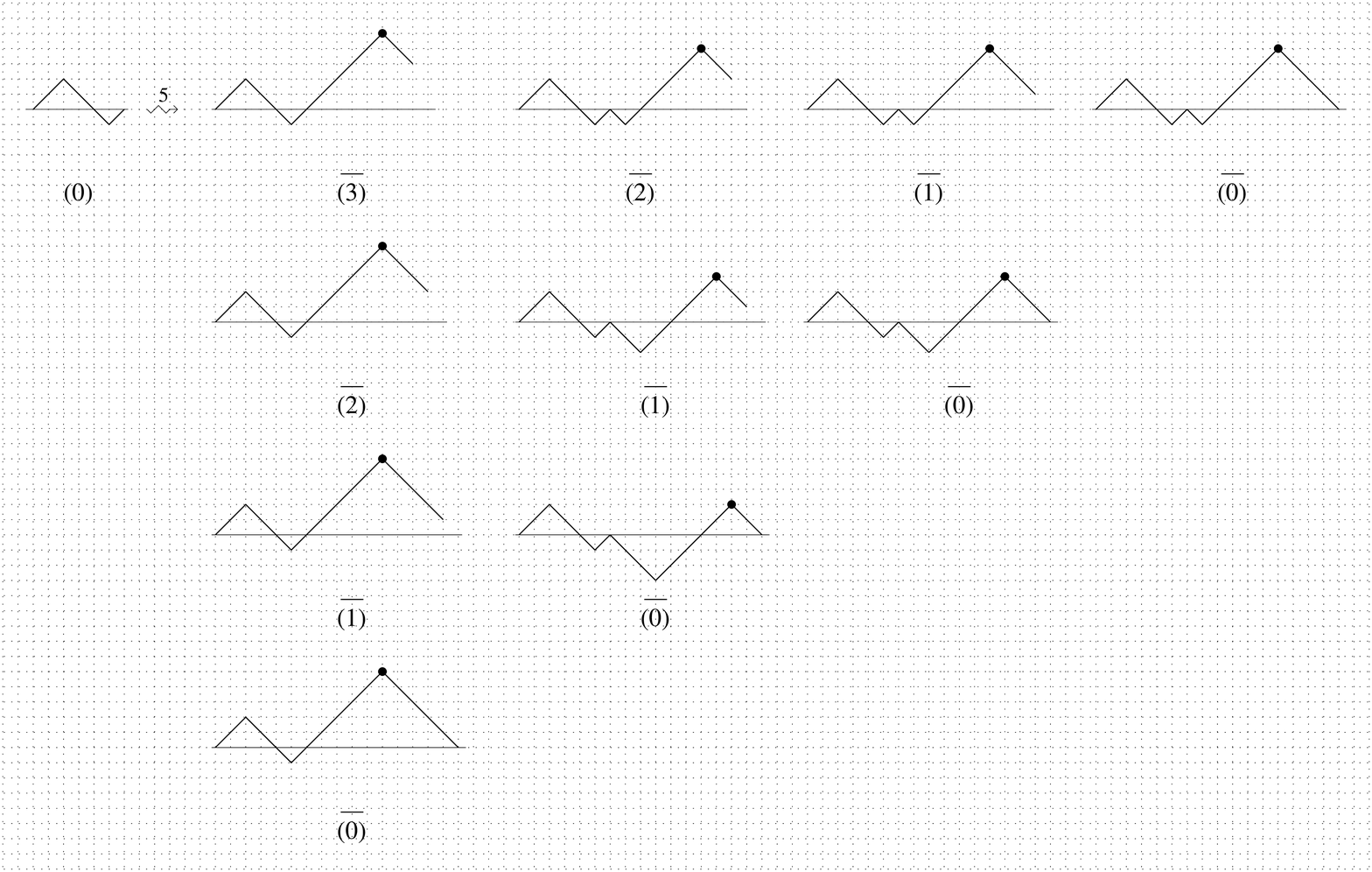,width=5.35in,clip=} \caption{\small{The
mapping associated to $(0) \stackrel{5}{\rightsquigarrow}
(\overline{0})^{4}
(\overline{1})^{3}(\overline{2})^{2}(\overline{3})$ on an
example}\label{zerata} }\vspace{-15pt}
\end{center}
\end{figure}

\item[$\bullet$] If the $\Delta$-path $\omega=\upsilon\rho$ in $F$
has the ordinate of its endpoint greater than 0, then we must have
to distinguish two cases: in the first one the rightmost suffix
$\rho$ in $\omega$ does not contain any marked points and in the
second one
$\rho$ contains at least one marked point.\\

\textbf{The suffix $\rho$ in $\omega$ does not contain any marked
point}. We denote by $h$ the ordinate of the peak in $\rho$ having
highest height. We consider the endpoint of the path $\omega x^j
\overline{x}^i$ having ordinate $k+j-i$ obtained from $\omega$ by
applying the mapping $(k) \stackrel{j}{\rightsquigarrow}
\overline{(k+j-i)}$ (see Figure \ref{sogliah}) and we distinguish
three subcases: $k+j-i \leq h$, $h < k+j-i < h+j-i$ and $k+j-i
\geq h+j-i$.

\begin{figure}[!htb]
\begin{center}
\epsfig{file=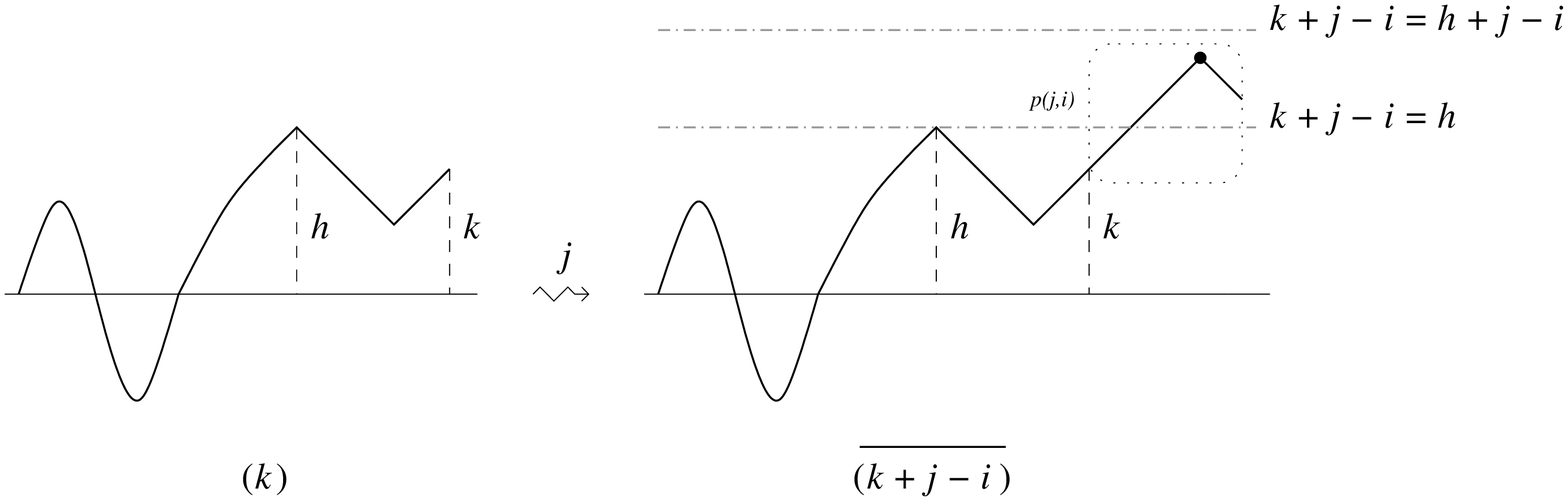,width=4.8in,clip=} \caption{\small{A
graphical representation of the path $\omega x^j \overline{x}^i$
when the suffix $\rho$ in $\omega$ does not contain any marked
point}\label{sogliah} }\vspace{-15pt}
\end{center}
\end{figure}

\begin{itemize}
\item[$\circ$] If $k+j-i \geq h+j-i$ (or equivalently $h-k \leq
0$), then $a=0$ and we apply to the path $\omega$ the production
(\ref{zero}) for the second production of (\ref{regola}) similarly
to the case $k=0$.

\begin{equation}
\label{zero} (k) \stackrel{j}{\rightsquigarrow}
(\overline{0})^{j-i+1}
(\overline{1})^{j-i}(\overline{2})^{j-i-1}\ldots
(\overline{j-i-1})^2(\overline{j-i})\dots(\overline{k+j-i})
\end{equation}

Note that, if the suffix $\rho$ in $\omega$ does not contain any
peak (or equivalently $h=0$) then we apply the production
(\ref{zero}) for the second production of (\ref{regola}). We can
observe that the paths $\omega''$ related to the previous
construction are the paths obtained from $\omega$ by adding a
sequence of steps consisting of the marked forbidden pattern
$\mathfrak{p}(j,i)=x^{j}\overline{x}^i$ followed by $y$ fall
steps, for any $y$ with $k \leq y \leq k+j-i-1$. In this case, the
point $z$ for the cut and paste actions is always the endpoint of
the path $\omega x^j \overline{x}^i$ having ordinate $k+j-i$.

\item[$\circ$] If $h < k+j-i < h+j-i$ (or equivalently
$0<h-k<j-i$), then $a=h-k$ and we apply to the path $\omega$ the
production (\ref{nelmezzo}) for the second production of
(\ref{regola}).

\begin{equation}
\label{nelmezzo} \begin{array}{lc}
(k)\stackrel{j}{\rightsquigarrow} (\overline{0})^{j-i+1-(h-k)}
(\overline{1})^{j-i-(h-k)}(\overline{2})^{j-i-1-(h-k)}\ldots\\\\
\ldots(\overline{j-i-1-(h-k)})^2(\overline{j-i-(h-k)})\dots(\overline{k+j-i})
\end{array}
\end{equation}

Also in this case, the point $z$ for the cut and paste actions is
always the endpoint of the path $\omega x^j \overline{x}^i$ having
ordinate $k+j-i$. The paths $\omega''$ related to the previous
construction are the paths obtained from $\omega$ by adding a
sequence consisting of the marked forbidden pattern
$\mathfrak{p}(j,i)=x^{j}\overline{x}^i$ followed by $y$ falls
steps, $h \leq y \leq k+j-i-1$.

\item[$\circ$] If $k+j-i \leq h$ (or equivalently $h-k \geq j-i$),
then $a=j-i-1$ and we apply to the path $\omega$ the production
(\ref{solouno}) for the second production of (\ref{regola}).

\begin{equation}
\label{solouno} (k) \stackrel{j}{\rightsquigarrow}
(\overline{0})^2
(\overline{1})(\overline{2})\dots(\overline{k+j-i})
\end{equation}

In this case the point $z$ for the cut and paste actions is always
the point of peak in $\rho$ having ordinate $h$, so we have only
one path $\omega''$ related to the previous construction, that is
the path obtained from $\omega$ by adding a sequence of steps
consisting of the marked forbidden pattern
$\mathfrak{p}(j,i)=x^{j}\overline{x}^i$ followed by $k+j-i-1$
falls steps.
\end{itemize}

\textbf{The suffix $\rho$ in $\omega$ contains at least one marked
point}. We denote by $h$ the ordinate of the no marked peak in
$\rho$ having highest height and by $h^*$ the ordinate of the
marked peak in $\rho$ having highest height. We consider the
endpoint of the path $\omega x^j \overline{x}^i$ having ordinate
$k+j-i$ obtained from $\omega$ by applying the mapping $(k)
\stackrel{j}{\rightsquigarrow} \overline{(k+j-i)}$ and we
distinguish three subcases: $h^*-h<i$ (see the left side of Figure
\ref{sogliahdue}), $h^*-h
> i$ (see the right side of Figure \ref{sogliahdue}) and $h^*-h=i$.

\begin{figure}[!htb]
\begin{center}
\epsfig{file=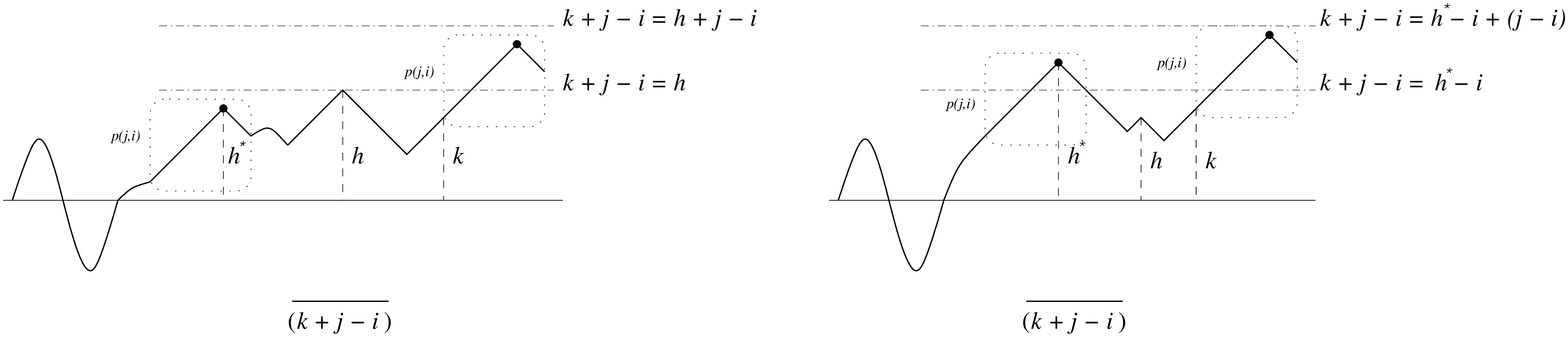,width=6.3in,clip=} \caption{\small{A
graphical representation of the path $\omega x^j \overline{x}^i$
when the suffix $\rho$ in $\omega$ contains at least one marked
point}\label{sogliahdue} }\vspace{-15pt}
\end{center}
\end{figure}
\begin{itemize}

\item[$\circ$] If $h^*-h<i$, just consider the no marked peak
having ordinate $h$ to set the parameter $a$, then we apply to the
path $\omega$ the productions related to the case in which the
suffix $\rho$ in $\omega$ does not contain any marked point.

\item[$\circ$] If $h^*-h > i$, then we consider three subcases:
$k+j-i \geq h^*-i+(j-i)$, $h^*-i<k+j-i<h^*-i+(j-i)$ and $k+j-i
\leq h^*-i$.

\begin{itemize}
\item[$\lozenge$] If $k+j-i \geq h^*-i+(j-i)$ (or equivalently
$h^*-h \leq i$), then $a=0$ and we apply to the path $\omega$ the
production (\ref{zero}) for the second production of
(\ref{regola}). We can observe that the paths $\omega''$ related
to the previous construction are the paths obtained from $\omega$
by adding a sequence of steps consisting of the marked forbidden
pattern $\mathfrak{p}(j,i)=x^{j}\overline{x}^i$ followed by $y$
fall steps, for any $y$ with $k \leq y \leq k+j-i-1$. In this case
the point $z$ for the cut and paste actions is always the endpoint
of the path $\omega x^j \overline{x}^i$ having ordinate $k+j-i$.

\item[$\lozenge$] If $h^*-i<k+j-i<h^*-i+(j-i)$ (or equivalently
$i<h^*-h<j$), then $a=h^*-k-i$ and we apply to the path $\omega$
the production (\ref{minc}) for the second production of
(\ref{regola}).

\begin{equation}
\label{minc}
\begin{array}{lc}
(k) \stackrel{j}{\rightsquigarrow}
(\overline{0})^{j-i+1-(h^*-k-i)}
(\overline{1})^{j-i-(h^*-k-i)}(\overline{2})^{j-i-1-(h^*-k-i)}\ldots\\\\
\ldots(\overline{j-i-1-(h^*-k-i)})^2
(\overline{j-i-(h^*-k-i)})\dots(\overline{k+j-i})
\end{array}
\end{equation}

Also in this case, the point $z$ for the cut and paste actions is
always the endpoint of the path $\omega x^j \overline{x}^i$ having
ordinate $k+j-i$. The paths $\omega''$ related to the previous
construction are the paths obtained from $\omega$ by adding a
sequence consisting of the marked forbidden pattern
$\mathfrak{p}(j,i)=x^{j}\overline{x}^i$ followed by $y$ falls
steps, $h^*-i \leq y \leq k+j-i-1$.

\item[$\lozenge$] If $k+j-i \leq h^*-i$ (or equivalently $h^*-h
\geq j$), then $a=j-i-1$ and we apply to the path $\omega$ the
production (\ref{solouno}) for the second production of
(\ref{regola}). In this case the point $z$ for the cut and paste
actions is always the first point, having ordinate $h^*-i$, on the
right of the marked forbidden pattern
$\mathfrak{p}(j,i)=x^{j}\overline{x}^i$ having ordinate $h^*$, so
we have only one path $\omega''$ related to the previous
construction, that is the path obtained from $\omega$ by adding a
sequence of steps consisting of the marked forbidden pattern
$\mathfrak{p}(j,i)=x^{j}\overline{x}^i$ followed by $k+j-i-1$
falls steps.
\end{itemize}

\item[$\circ$] If $h^*-h=i$ and the no marked peak having ordinate
$h$ is on the left of the marked peak having ordinate $h^*$ then
we apply to the path $\omega$ the productions related to the case
in which the suffix $\rho$ in $\omega$ does not contain any marked
point, otherwise if $h^*-h=i$ and the no marked peak having
ordinate $h$ is on the right of the marked peak having ordinate
$h^*$ then we apply to the path $\omega$ the productions related
to the case $h^*-h > i$.
\end{itemize}
\end{itemize}

\subsection{$\Gamma$-paths in $F$}
For each $\Gamma$-path $\omega$ in $F$ having $k$ as ordinate of
its endpoint, we apply the following succession rule
(\ref{regolabrutta}), for each $k \geq 1$:
\begin{equation}
\label{regolabrutta} \left\{
\begin{array}{cl}
 (k) & \stackrel{1}{\rightsquigarrow} (0) (1) (2)\cdots (k)(k+1)\\
 (k) & \stackrel{j}{\rightsquigarrow} (\overline{0}) (\overline{1})(\overline{2})
  \cdots (\overline{k+j-i-1})(\overline{k+j-i})
\end{array}
\right.
\end{equation}

A $\Gamma$-path $\omega \in F$, with $n$ rise steps and such that
its endpoint has ordinate $k$, provides $k+2$ lattice paths, with
$n+1$ rise steps, according to the first production of
(\ref{regolabrutta}) having $0,1,\ldots,k+1$ as endpoint ordinate,
respectively. Those labels are obtained by adding to $\omega$ a
sequence of steps consisting of one rise step followed by $k+1-y$,
$0 \leq y \leq k+1$, fall steps.

Moreover, a $\Gamma$-path $\omega \in F$, with $n$ rise steps and
such that its endpoint has ordinate $k$, provides $1+k+j-i$
lattice paths, with $n+j$ rise steps, according to the second
production of (\ref{regolabrutta}) having $0,1,2,\ldots,k+j-i$ as
endpoint ordinate, respectively. Those labels are obtained by
adding to $\omega$ a sequence of steps consisting of the marked
forbidden pattern $\mathfrak{p}=x^{j}\overline{x}^i$ followed by
$k+j-i-y$ fall steps, $0 \leq y \leq k+j-i$.\\

The just described construction, both for $\Delta$-paths and
$\Gamma$-paths in $F$, generates $2^C$ copies, $C \geq 0$, of each
path having $C$ forbidden patterns such that $2^{C-1}$ are coded
by a sequence of labels ending by a marked label, say
$(\overline{k})$, and contain an odd number of marked forbidden
pattern, and $2^{C-1}$ are coded by a sequence of labels ending by
a non-marked label, say $(k)$, and contain an even number of
marked forbidden pattern.

For brevity sake, we omit the proof of the fact that the described
algorithm is a construction for $F^{[\mathfrak{p}(j,i)]}$, where
$\mathfrak{p}(j,i)=x^{j}\overline{x}^i=1^{j}0^i$, $0<i<j$. In
order to prove the theorem we should have to show that the
described actions are uniquely invertible.

\section{Conclusions and further developments}
In this paper we propose an algorithm for the construction,
according to the number of ones, of particular binary words
excluding a fixed pattern $\mathfrak{p}(j,i)=1^{j}0^i$, $0<i<j$.

Successive studies should take into consideration binary words
avoiding different forbidden patterns both from an enumerative and
a constructive point of view.

Afterwards, it should be interesting to study words avoiding
patterns having a different shape, that is not only patterns
consisting of a sequence of rise steps followed by a sequence of
fall steps. This could be the first step to investigate a possible
uniform generating algorithm for pattern avoiding words.

One could also consider a forbidden pattern on an arbitrary
alphabet and investigating words avoiding that pattern.

Finally, we could think to study words avoiding more than one
pattern and the related combinatorial objects, considering various
parameters.

\end{document}